\documentclass[11pt]{article}
\usepackage{amssymb}
\usepackage{graphicx}
\usepackage{pstricks}
\usepackage[rflt]{floatflt}
\newcommand{\ve}{\mathbf}
\newcommand{\x}{\theta}
\newcommand{\equal}{\!\!\!&=&\!\!\!}

\begin{document}
\abovedisplayshortskip 12pt
\belowdisplayshortskip 12pt
\abovedisplayskip 12pt
\belowdisplayskip 12pt
\baselineskip=15pt
\title{{\bf Integrability of a nonlinear evolution equation derived from isoperimetric plane curve motion}}
\author{J. C. Brunelli\thanks{\texttt{jcbrunelli@gmail.com}}  \\
\\
Departamento de F\'\i sica, CFM\\
Universidade Federal de Santa Catarina\\
Campus Universit\'{a}rio, Trindade, C.P. 476\\
CEP 88040-900\\
Florian\'{o}polis, SC, Brazil\\
}
\date{}
\maketitle

\begin{center}
{ \bf Abstract}
\end{center}

We provide a geometrical interpretation for the series of transformations used by Sakovich to map the third-order nonlinear evolution equation obtained by Chou and Qu to the mKdV equation. We also discuss its bi-Hamiltonian integrability as well as integrable equations associated with this system.
\bigskip



\newpage

\section{Introduction:}

Starting with the work of Hasimoto \cite{Hasimoto} the study of the connection between integrable models and the dynamics of space curves has drawn a lot of interest. Motion of these curves and the associated integrable models were considered in Euclidean space $E^3$ \cite{Lamb,Lakshmanan,LangerPerline}, $S^2$ and $S^3$ spaces \cite{DoliwaSantini} and Minkowski space \cite{Nakayama,Gurses}. However, integrable equations associated with the motion of plane curves are more scarce \cite{GoldsteinPetrich,NakayamaSegurWadati}. This has motivated Chou and Qu \cite{ChouQu1,ChouQu2} to study motions of plane curves in other background geometries, replacing the Euclidean geometries by Klein ones. As a result they have shown that KdV, Harry-Dym and Sawada-Kotera hierachies, among other nonlinear equations,  arise in an natural way from the motions of plane curves in affine, centro-affine and similarity geometries. In this process they also found a new equation associated with the mKdV equation, namely (see \cite{ChouQu1}, last equation on page 31),
\begin{equation}
u_t={1\over 2}\Bigl((u_{xx}+u)^{-2}\Bigr)_x\;.\label{theequation}
\end{equation}
Chou and Qu did not investigated  if this equation (from now on the CQ equation) arises from an AKNS- or the WKI-scheme of inverse scattering transformation, its integrability was investigated by Sakovich \cite{Sakovich}. Through a chain of Miura-type transformations he related equation (\ref{theequation}) with the mKdV equation. Firstly, (\ref{theequation}) can be transformed to
\begin{equation}
v_t={1\over2}v^2\Bigl((v^2)_{xx}+v^2\Bigr)_x\label{vtimex}\;,
\end{equation}
by the transformation
\begin{equation}
(x,t,u(x,t))\mapsto(x,t,v(x,t)):\;v=-(u_{xx}+u)^{-1}\label{utov}\;.
\end{equation}
Then we perform two successive transformations
\begin{eqnarray}
(y,t,w(y,t))&\!\!\mapsto\!\!&(x,t,v(x,t)):\;x=w\;,\;\;v=w_y\;,\nonumber\\\noalign{\vskip -1pt}
(y,t,w(y,t))&\!\!\mapsto\!\!& (y,t,z(y,t)):\;z=w_y\label{wtoy}\;,
\end{eqnarray}
to obtain the mKdV equation
\begin{equation}
z_t=z_{yyy}+{3\over2}z^2z_y\label{zmkdv}\;,
\end{equation}
where $z(y,t)=v(x,t)$. Also, in \cite{Sakovich} a zero curvature representation with an essential parameter was obtained as well as the following second-order recursion operator
\begin{equation}
{\overline R}={1\over u_{xx}+u}\partial{1\over u_{xx}+u}(\partial+\partial^{-1})\;.\label{urecursion}
\end{equation}

It is well known that integrable equations also possess a bi-Hamiltonian structure which yields a recursion operator in the form ${\cal D}_2{\cal D}_1^{-1}$. So, going backwards Ren and Alatancang \cite{RenAlatancang} proved the hereditary property of (\ref{urecursion}) and by a decomposition of it obtained the bi-Hamiltonian structure as well as a hierarchy of equations, associated with (\ref{theequation}).

In this paper we point out, in Section 2, that the series of transformations (\ref{utov}) and (\ref{wtoy}) mapping the CQ equation (\ref{theequation}) to the mKdV equation (\ref{zmkdv}) has a nice geometrical interpretation which can be used to justify connections between other equations. In Section 3 the bi-Hamiltonian structure, and consequently the recursion operator $R$, is derived directly from a Lagrangian representation of (\ref{theequation}). In fact, the original derivation of Ren and Alatancang  for this bi-Hamiltonian structure relies on the knowledge of the bi-Hamiltonian structure of a system related to (\ref{theequation}) by a Miura map (obtained previously by Olver and Rosenau \cite{OlverRosenau}) and of the recursion operator (\ref{urecursion}). Therefore, their derivation is more a map between structures than a derivation from first principles. For completeness, in Section 4, we obtain the hierarchy of equations, some symmetries and equations related to (\ref{theequation}). We discuss the relation among these equations. The conclusions are presented in Section 5.

\section{Plane Curve Motion:}

Let  be a closed smooth curve in the plane, parametrized by an arbitrary parameter $\alpha$, where $\alpha\in[0,1]$, and let  ${\ve r}(\alpha,t)$ represent the position of a point on the curve at the time $t$.
Along the curve the arc-length is $s(\alpha,t)=\int_0^\alpha\sqrt{g(\alpha',t)}\, d\alpha'$ and we can use $s$ as a parameter as well (this is called Lagrangian description of the curve). The metric on the curve is $g={\ve r}_\alpha\cdot{\ve r}_\alpha$ and we have that $d/d\alpha=\sqrt{g}\,d/ds$. At a point on the curve the unit tangent and normal vectors, defined by ${\ve t}(s,t)={\ve r}_s$ and ${\ve n}(s,t)={1\over\kappa}{\ve r}_{ss}$, respectively, satisfy the Serret-Frenet equations
\begin{equation}
\left(%
\begin{array}{c}
  {\ve t} \\
  {\ve n} \\
\end{array}%
\right)_s
=
\left(%
\begin{array}{cc}
  0 & \kappa \\
  -\kappa & 0 \\
\end{array}%
\right)
\left(%
\begin{array}{c}
  {\ve t} \\
  {\ve n} \\
\end{array}%
\right)
=i\sigma_2\kappa
\left(%
\begin{array}{c}
  {\ve t} \\
  {\ve n} \\
\end{array}%
\right)
\;,\label{tns}
\end{equation}
where $\kappa(s,t)$ is the curvature. The dynamics of the points of the curve are specified by
\begin{equation}
{\ve r}_t=F{\ve n}+ G{\ve t}\label{evol}\;,
\end{equation}
where the normal and tangential velocities $F$ and $G$ are functions of the curvature $\kappa$. Now we make the further assumption that the perimeter $L=\oint ds$ of the closed curve remains constant in time, i.e., we assume an isoperimetric plane curve motion. Therefore, from the metric evolution
\[
g_t=2{\ve r}_\alpha\cdot{\ve r}_{t\alpha}=2g\left(G_s-\kappa F\right)\;,
\]
we get
\[
{d L\over d t}=\int_0^1\sqrt{g(\alpha',t)}\, d\alpha'=\oint\left(G_s-\kappa F\right)ds\;,
\]
and ${d L/d t}=0$ if $G_s=\kappa F$. This implies that the arc-length and time derivatives commute, $[\partial_s,\partial_t]=0$. The time evolution of the tangent and normal vectors gives
\begin{equation}
\left(%
\begin{array}{c}
  {\ve t} \\
  {\ve n} \\
\end{array}%
\right)_t
=
\left(%
\begin{array}{cc}
  0 & F_s+\kappa G \\
  -F_s-\kappa G& 0 \\
\end{array}%
\right)
\left(%
\begin{array}{c}
  {\ve t} \\
  {\ve n} \\
\end{array}%
\right)
=i\sigma_2(F_s+\kappa G)
\left(%
\begin{array}{c}
  {\ve t} \\
  {\ve n} \\
\end{array}%
\right)
\;,\label{tnt}
\end{equation}
and the compatibility condition between (\ref{tns}) and (\ref{tnt}),
\[
\left(%
\begin{array}{c}
  {\ve t} \\
  {\ve n} \\
\end{array}%
\right)_{ts}=\left(%
\begin{array}{c}
  {\ve t} \\
  {\ve n} \\
\end{array}%
\right)_{st}\;,
\]
yields the following nonlinear evolution equation for the curvature
\begin{equation}
\kappa_t=F_{ss}+\kappa_s\,\partial^{-1}_s(\kappa F) + \kappa^2 F\label{kevol}\;,
\end{equation}
which can also be written as $\kappa_t=RF$ with
\begin{equation}
R=\partial_s^2+\kappa_s\partial^{-1}_s\kappa+\kappa^2\;.\label{srecursion}
\end{equation}
\begin{center}
\includegraphics[height=.4\hsize]{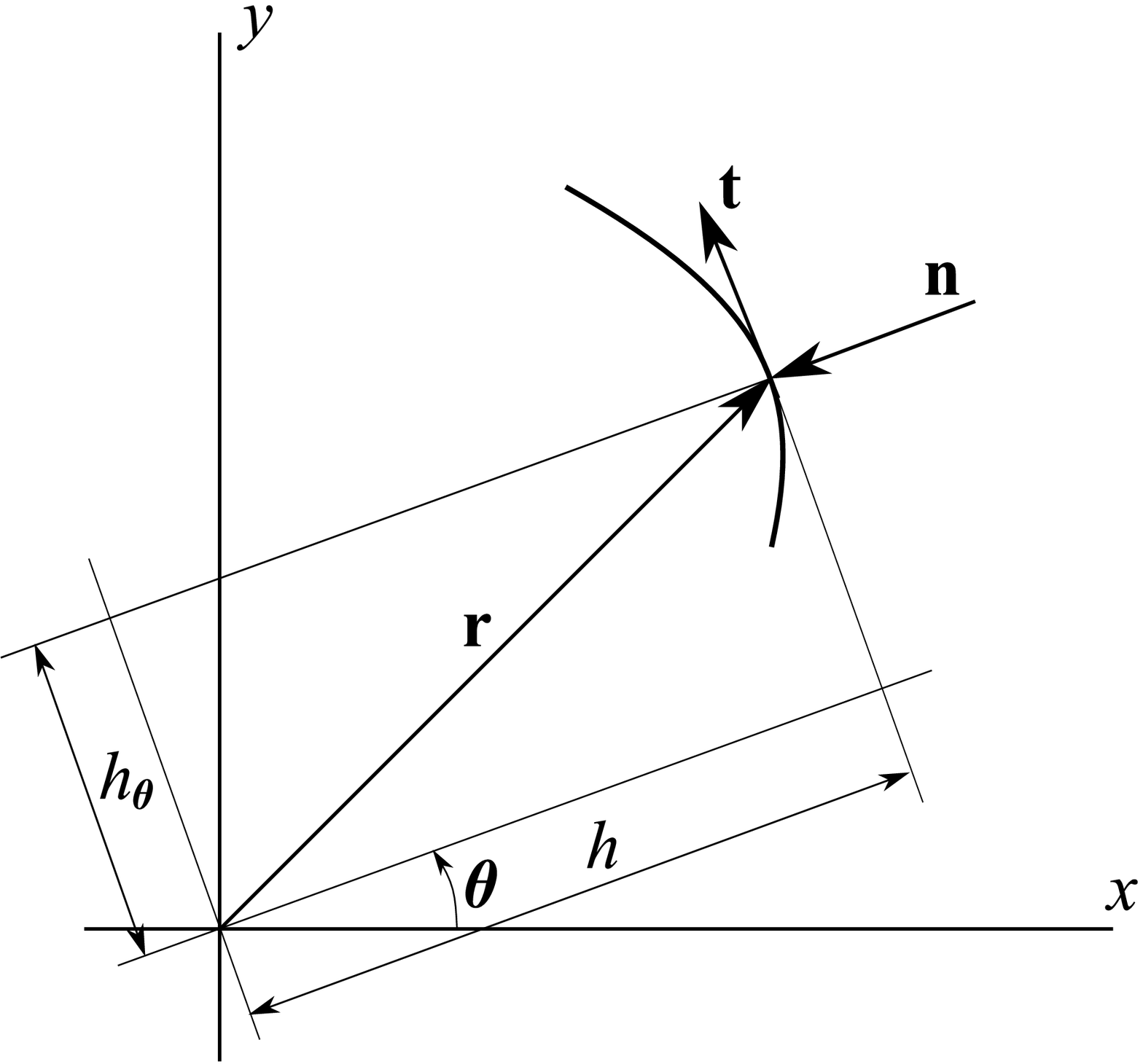}
\centerline{Figure 1}
\end{center}
If we choose $F=-\kappa_s$ in (\ref{kevol}) we obtain the mKdV equation,
\begin{equation}
\kappa_t=-\kappa_{sss}-{3\over2}\kappa^2\kappa_s\;,\label{mkdv}
\end{equation}
and we recognize (\ref{srecursion}) as the mKdV recursion operator. However, at this point let us change the curve description and parametrization. For a curve with positive curvature (uniformly convex) we introduce the support function $h(\x,t)$, defined as
\begin{equation}
h(\x,t)=-{\ve r}\cdot{\ve n}\label{support}\;,
\end{equation}
where  $\x$, called the normal angle of the curve, is the angle between the unit normal $\ve n$ and the positive $x$-axis \cite{Guggenheimer}. In the following let ${\ve r}={\ve r}(\x,t)$ and from the Figure 1 we have ${\ve n}=-(\cos\x,\sin\x)$ and ${\ve t}=(-\sin\x,\cos\x)$. For the support function $h(\x,t)$ we obtain
\begin{eqnarray}
h\equal x\cos\x+y\sin\x\nonumber\\\noalign{\vskip -1pt}
h_\x\equal -x\sin\x+y\cos\x\nonumber\;,
\end{eqnarray}
and ${\ve r}=(x,y)$ can be obtained form $h$ using
\begin{eqnarray}
x\equal h\cos\x-h_\x\sin\x\nonumber\\\noalign{\vskip -1pt}
y\equal h\sin\x+h_\x\cos\x\nonumber\;.
\end{eqnarray}
From this last equation we obtain
\[
|{\ve r}_\x|=h_{\x\x}+h\;,
\]
and since $d/ds=|{\ve r}_\x|^{-1}d/d\x$ the Serrat-Frenet formula
\[
{d{\ve t}\over ds}=|{\ve r}_\x|^{-1}{d{\ve t}\over d\x}=\kappa{\ve n}
\]
yields a neat expression for the curvature in terms of the support function
\begin{equation}
\kappa={1\over{h_{\x\x}+h}}\label{curvatureh}\;.
\end{equation}
Finally, from (\ref{evol}) the normal speed of $\ve r$ is given by ${\ve n}\cdot{\ve r}_t=-\kappa_s$ and therefore
\begin{eqnarray}
h_t\equal\kappa_s=|{\ve r}_\x|^{-1}\kappa_\x={1\over2}(\kappa^2)_\x\nonumber\\\noalign{\vskip -1pt}
\equal{1\over 2}\left((h_{\x\x}+h)^{-2}\right)_\x\label{htks}\;,
\end{eqnarray}
which is the equation (\ref{theequation}).
An evolution equation for the curvature, in terms of the normal angle $\x$, instead of the arc-length as in (\ref{mkdv}), can be obtained. From (\ref{curvatureh})
\[
\kappa_t=-\kappa^2(h_{\x\x t}+h_t)\;,
\]
and after using (\ref{htks})
\begin{equation}
\kappa_t=-{1\over2}\kappa^2\Bigl((\kappa^2)_{\x\x}+\kappa^2\Bigr)_\x\;.\label{kappatimetheta}
\end{equation}
It is well know that a curve is determined by its curvature up to a rigid motion \cite{Guggenheimer}, therefore, there is a formal equivalence between (\ref{kappatimetheta}) and (\ref{htks}) (and between (\ref{evol}) and (\ref{mkdv})) and the connection is made by (\ref{curvatureh}). In other words the equation (\ref{htks}) and the curvature integrable equation (\ref{kappatimetheta}) are equivalent. Now we are in position to justify geometrically the transformations (\ref{utov}) and  (\ref{wtoy}) if we make the identifications $h(\x,t)\leftrightarrow u(x,t)$, $\kappa(s,t)\leftrightarrow z(y,t)$, $\kappa(\x,t)\leftrightarrow v(x,t)$, $\x\leftrightarrow x$ and $s\leftrightarrow y$ and recognize the equivalence between the equations $(\ref{htks})\leftrightarrow(\ref{theequation})$, $(\ref{mkdv})\leftrightarrow(\ref{zmkdv})$ and $(\ref{kappatimetheta})\leftrightarrow(\ref{vtimex})$. The transformation (\ref{utov}) is simply (\ref{curvatureh}) and the transformation (\ref{wtoy}) is the inverse transformation from (\ref{mkdv}) to (\ref{kappatimetheta}) (projection (\ref{support}) and change of parametrization from $s$ to $\x$). The recursion operator (\ref{urecursion}) defines the curvature evolution under arc-length conserving dynamics and can be obtained from (\ref{srecursion}) using the identifications given above.
\section{Bi-Hamiltonian Structure:}

Let us study the integrability of ({\ref{htks}) from a Hamiltonian point of view. Since  the curvature  plays a prominent role let us write ({\ref{htks}) as
\begin{equation}
h_t={1\over 2}(\kappa^2)_\x\;,\label{kappaequation}
\end{equation}
and where $\kappa$ satisfies (\ref{kappatimetheta}) which can also be rewritten as
\begin{equation}
(\kappa^{-1})_t={1\over2}\Bigl((\kappa^2)_{\x\x}+\kappa^2\Bigr)_\x\;.\label{1overkappatimetheta}
\end{equation}
Let us observe that the basic field is $h$ and that $\kappa$ is a placeholder used to make expressions more compact (see \cite{Brunelli}). The equations (\ref{kappatimetheta}) and (\ref{1overkappatimetheta}) are not being viewed as nonlinear evolution equations but just as expressions for the time derivative of the placeholder $\kappa$, however, they can be interpreted as the Casimir equation for the so called modified compacton hierarchy (see equations (27) and (28) in \cite{OlverRosenau}). Also, we will consider $\x$-periodic solutions, therefore, $\x\in[0,2\pi]$ and $h(\x,t)=h(\x+2\pi,t)$. We denote by $\partial\equiv\partial_\x$ the differential operator with respect to $\x$ with skew-adjoint inverse $\partial^{-1}\equiv\partial^{-1}_\x$. Whenever the nonlocal operator $\partial^{-1}$ is used we have $\partial\partial^{-1}=\partial^{-1}\partial=I$, $\partial^\dagger=-\partial$ and $\left(\partial^{-1}\right)^\dagger=-\partial^{-1}$. We define the anti-derivative $\partial^{-1}$ as $\int^\x_0\,d\x'$.

Equation ({\ref{kappaequation}) can be obtained from a variational principle, $\delta\int dtd\x\,{\cal L}$, from the Lagrangian density
\begin{equation}
{\cal L}=\kappa-h_t\,\partial^{-1}\kappa^{-1}\;.\label{Lagrangian}
\end{equation}
This is a first order Lagrangian density and we can use the
Dirac's theory of constraints \cite{Dirac} to obtain the
Hamiltonian and the Hamiltonian operator associated with
(\ref{Lagrangian}). The Lagrangian is degenerate and the primary
constraint is obtained to be
\begin{equation}
\Phi=\pi+h_\x+(\partial^{-1}h)\;,\label{primary}
\end{equation}
where $\pi={{\partial{\cal L}}/{\partial h_t}}$ is the canonical
momentum. The total Hamiltonian is
\begin{equation}
H_T =\int d\x\left(\pi h_t-{\cal L}+\lambda\Phi\right)
=\int d\x\left[ -\kappa  +\lambda\left(\pi+h_\x+(\partial^{-1}h)\right)\right]\;,\label{ht}
\end{equation}
where $\lambda$ is a Lagrange multiplier field. From the canonical
Poisson bracket relation
\begin{equation}
\{h(\x),\pi(\x')\}=\delta(\x-\x')\;,\label{poisson}
\end{equation}
with all others vanishing, we require the
primary constraint to be stationary under time evolution,
\[
\{\Phi(\x),H_T\}=0\;,
\]
to determine the Lagrange multiplier field $\lambda$ in (\ref{ht}) and to find out that
the system has no further constraints.

The canonical Poisson bracket relation (\ref{poisson}) yields
\begin{equation}
K(\x,\x')\equiv\{\Phi(\x),\Phi(\x')\}=\left( \partial_{\x}+{\partial_{\x}^{-1}}\right) \delta(\x-\x')-\left(\partial_{\x'}+{\partial_{\x'}}^{-1}\right)\delta(\x'-\x)\;,\label{kpoisson}
\end{equation}
and we find that the constraint (\ref{primary}) is second class. The Dirac bracket between the basic variables is
\[
\{h(\x),h(\x')\}_D=\{h(\x),h(\x')\}-\int
d\x_1\,d\x_2\{h(\x),\Phi(\x_1)\}J(\x_1,\x_2)\{\Phi(\x_2),h(\x')\}=J(\x,\x')\;,\
\]
where $J$ is the inverse of the Poisson bracket of the constraint
(\ref{kpoisson}),
\[
\int d\x''\,K(\x,\x'') J(\x'',\x')=\delta(\x-\x')\,.
\]
From this last equation we get
\[
2\left( \partial+{\partial^{-1}}\right)  J(\x,\x')=\delta(\x-\x')\;,
\]
or
\[
J(\x,\x')={\cal D}_1\delta(\x-\x')\;,
\]
where
\begin{equation}
{\cal D}_1={1\over 2}\left( \partial+{\partial^{-1}}\right)^{-1}\;.\label{d1}
\end{equation}
We now set the constraint (\ref{primary})
strongly to zero in (\ref{ht}) to obtain
\begin{equation}
H_2 =-\int d\x\kappa\;,\label{h2}
\end{equation}
and the  equation (\ref{kappaequation}) can be written in the Hamiltonian form as
\[
h_t={\cal D}_1{\delta H_2\over\delta h}\;.
\]

From (\ref{1overkappatimetheta}) it is straightforward to show that the charge
\begin{equation}
H_1 = \int d\x {1\over \kappa}\label{h1}
\end{equation}
is also conserved. Therefore,  the equation (\ref{kappaequation}) can be written in the Hamiltonian form as
\[
h_t={\cal D}_2{\delta H_1\over\delta h}\;,
\]
where we have defined
\begin{equation}
{\cal D}_2=\kappa\partial\kappa\;.\label{d2}
\end{equation}
This Hamiltoninan structure is manifestly skew symmetric. Using the expansion
\begin{equation}
\left(\partial+\partial^{-1}\right)^{-1}=\sum_{n=0}^\infty(-1)^n\partial^{-(2n+1)}\;\label{expansion}
\end{equation}
it is easy to show that ${\cal D}_1$ is also skew symmetric. Jacobi identity for these
structures as well as their compatibility  follows from standard method of
prolongation \cite{Olver}. We can construct the two bivectors associated with the two
structures as
\begin{eqnarray}
\Phi_{{\cal D}_1}\equal{1\over 2}\int d\x\,\left\{\phi\wedge{\cal
D}_1\phi\right\}={1\over4}\int d\x\,\phi\wedge\left[\left(\partial+\partial^{-1}\right)^{-1}\phi\right]
\;,\nonumber\\\noalign{\vskip 5pt}
\Phi_{{\cal D}_2}\equal{1\over 2}\int d\x\,\left\{\phi\wedge{\cal
D}_2\phi\right\}={1\over2}\int
d\x\,\kappa^2\phi\wedge\phi_\x\;.\nonumber
\end{eqnarray}
where $\phi$ is the univector corresponding to the one-form $dh$.
Using the prolongation relations,
\begin{eqnarray}
\hbox{\bf pr}\,{\vec v}_{{\cal D}_1{\phi}} (\kappa^2) \equal -\kappa^3\phi_\x \nonumber\;,\\ \noalign{\vskip 5pt}
\hbox{\bf pr}\,{\vec v}_{{\cal D}_2{\phi}} (\kappa^2) \equal -2\kappa^3\left[\left(\partial+\partial^{-1}\right)\kappa(\kappa\phi)_\x\right]_\x\;,\label{prolongation}
\end{eqnarray}
we show that the prolongation of the
bivector $\Phi_{{\cal D}_2}$ vanishes,
\[
\hbox{\bf pr}\,{\vec v}_{{\cal D}_2\phi}\left(\Phi_{{\cal
D}_2}\right)=0\;,
\]
implying that ${\cal D}_2$ satisfies  Jacobi identity.  Using
(\ref{prolongation}), it also follows that
\[
\hbox{\bf pr}\,{\vec v}_{{\cal D}_1\phi}\left(\Phi_{{\cal
D}_2}\right)+\hbox{\bf pr}\,{\vec v}_{{\cal
D}_2\phi}\left(\Phi_{{\cal D}_1}\right)=0\;,
\]
showing that ${\cal D}_1$ and ${\cal D}_2$ are compatible. Namely,
not only are ${\cal D}_{1}, {\cal D}_{2}$ genuine Hamiltonian
structures, any arbitrary linear combination of them is as well.
As a result, the dynamical equation (\ref{kappaequation}) is bi-Hamiltonian and, consequently, is integrable \cite{Olver,Magri}. Also, the bi-Hamiltonian structures (\ref{d1}) and (\ref{d2}) provide us with a natural recursion operator defined by
\[
R={\cal D}_2{\cal D}_1^{-1}\;\label{r}\;,
\]
or
\[
R=2\kappa\partial\kappa( \partial+{\partial^{-1}})\;,
\]
which is exactly the recursion operator $\overline R$, given by (\ref{urecursion}), obtained by Sakovich \cite{Sakovich}.

\section{Hierarchy of Equations:}

From the bi-Hamiltonian structure obtained in the previous section we can naturally define a hierarchy of commuting flows by
\begin{equation}
h_{t_n}=K_n[h]={\cal D}_1{\delta H_{n+1}\over\delta h}={\cal
D}_2{\delta H_{n}\over\delta h}\;,\quad
n=1,2,3,\dots\;.\label{eqrecursion}
\end{equation}
For $n=1$ we obtain (\ref{kappaequation}). From (\ref{eqrecursion}) we have
\begin{equation}
{\delta H_{n+1}\over\delta h}=R^\dagger{\delta H_{n}\over\delta
h}\;,\label{hrecursion}
\end{equation}
where
\[
R^\dagger=({\cal D}_2{\cal D}_1^{-1})^\dagger=2(\partial^{-1}+\partial)\kappa\partial\kappa\label{rdagger}
\]
is the adjoint of $R$. Using (\ref{hrecursion}) recursively we obtain an infinite set of conserved Hamiltonians
\begin{eqnarray}
H_1\equal\int d\x\,{1\over\kappa}\;,\nonumber\\\noalign{\vskip 7pt}
H_2\equal-\int d\x\,\kappa\;,\nonumber\\\noalign{\vskip 7pt}
H_3\equal-{1\over2}\int d\x\left(\kappa^3-4\kappa\kappa_\x^2\right)\;,\nonumber\\\noalign{\vskip 7pt}
H_4\equal-24\int d\x\,\left(3\kappa^5+20\kappa^3\kappa_{\x\x}^2+15\kappa^4\kappa_{\x\x}+
\kappa^4\kappa_{\x\x\x\x}\right)\;,\nonumber\\\noalign{\vskip 7pt}
H_5\equal-2\int  d\x\,\left[{\frac{5}{16}}\kappa^7+{\frac{35}{12}}\kappa^6\kappa_{\x\x}+
7\left({\frac{13}{120}}\kappa^6\kappa_{\x\x\x\x}+
{\frac{27}{20}}\kappa^5\kappa_{\x\x}^2\right)\right.\nonumber\\\noalign{\vskip 7pt}
&-&\!\!\!\left.
{\frac{28}{15}}\kappa^5\kappa_{\x\x\x}^2
-{\frac{21}{2}}\kappa^4\kappa_\x^2\kappa_{\x\x\x\x}
+{\frac{21}{2}}\kappa^4\kappa_{\x\x}^3
+{\frac{16}{45}}\kappa^6\kappa_{\x\x\x\x\x\x}
\right]\;,\nonumber\\
\noalign{\vskip 7pt}
&\,\,\,\vdots\;.\label{local}
\end{eqnarray}
These charges can also be obtained using the fact that
\[
H_{n}=\hbox{Tr}R^{2n-3\over2}\;,\quad
n=1,2,3,\dots\;,
\]
where ``Tr'' is the Adler's trace \cite{Adler}. We also obtained (\ref{local}) in this way using our program PSEUDO \cite{Brunelli1}.

The corresponding first flows associated with (\ref{local}) are
\begin{eqnarray}
h_{t_{1}}\equal {1\over2}\left(\kappa^2\right)_\x\;,\nonumber\\\noalign{\vskip 7pt}
h_{t_{2}}\equal\displaystyle\left({3\over4}\kappa^2+2\kappa^3\kappa_{\x\x}+\kappa^2\kappa_\x^2\right)_{\x}\;,\nonumber\\\noalign{\vskip 7pt}
h_{t_{3}}\equal \displaystyle\left({5\over4}\kappa^6+15\kappa^4\kappa_\x^2
+10\kappa^5\kappa_{\x\x}
+18\kappa^4\kappa_{\x\x}^2
+24\kappa^4\kappa_\x\kappa_{\x\x\x}
+4\kappa^5\kappa_{\x\x\x\x}
+2\kappa^2\kappa_{\x}^4
+32\kappa^3\kappa_{\x}^2\kappa_{\x\x}
\right)_{\x}\;,\nonumber\\
\noalign{\vskip 7pt}
&\,\,\,\vdots\;.
\label{localflow}
\end{eqnarray}

Since every symmetry of an integrable model defines another integrable model we can go further in the study of the hierarchy of equations of our system through a symmetry study of Eq. (\ref{kappaequation}). Using the Lie's algorithm (assisted by the computer algebra system program GeM \cite{Cheviakov}) we have obtained the following point and first higher order symmetry generators in evolutionary form
\begin{eqnarray}
X_1\equal h_\x{\partial\ \over\partial h}\;,\nonumber\\\noalign{\vskip 7pt}
X_2\equal \sin\x{\partial\ \over\partial h}\;,\nonumber\\\noalign{\vskip 7pt}
X_3\equal \cos\x{\partial\ \over\partial h}\;,\nonumber\\\noalign{\vskip 7pt}
X_4\equal -{1\over2}\left(\kappa^2\right)_\x{\partial\ \over\partial h}\;,\nonumber\\\noalign{\vskip 7pt}
X_5\equal \left[-{t\over2}\left(\kappa^2\right)_\x+{h\over3}\right]{\partial\ \over\partial h}\;.\nonumber
\end{eqnarray}
Of course the flow associated with $X_4$ is in the hierarchy  (\ref{localflow}). The flow associated with $X_2$ can be obtained if we allow in (\ref{eqrecursion}) the value $n=0$. In fact, from
\[
h_{t_0}={\cal D}_1{\delta H_{1}\over\delta h}\;,
\]
where ${\delta H_{1}/\delta h}=1$, and after using (\ref{expansion}) and $(\partial^{-1}1)={\theta^n/n!}$ we obtain
\[
h_{t_0}={1\over2}\left(\partial+\partial^{-1}\right)^{-1}\cdot1={1\over2}\sum_{n=0}^\infty(-1)^n\left(\partial^{-(2n+1)}1\right)={1\over2}\sin \theta\;.
\]
 From the flow $X_1$ we take $h_t=h_\theta$ as a seed equation and  the recursion procedure $h_t=(R^{-1})^n h_\theta$ generates for $n=1$
\[
h_t+h_{\theta\theta t}={1\over4}\left[(h+h_{\theta\theta})(h^2+h^2_\theta)\right]_\theta\;.
\]
This is the dual counterpart of the mKdV equation considered by Olver and Rosenau (equation (25) in \cite{OlverRosenau}). In this same work we recognize (up to multiplicative constants) equation (\ref{kappatimetheta}) as the Lagrange transform of the mKdV equation (equation (28) in \cite{OlverRosenau}). From (\ref{kappatimetheta}) (or (\ref{1overkappatimetheta})) if we make the identification $\kappa=1/\rho$ we obtain
\begin{equation}
\rho_t={1\over2}(\rho^{-2})_{\x\x\x}+{1\over2}(\rho^{-2})_\x\;,\label{mkdvcasimir}
\end{equation}
called Casimir equation for the modified compacton hierarchy (up to multiplicative constants it is the equation (27) in \cite{OlverRosenau}). It is implicit in \cite{OlverRosenau} (due to the role played by the mKdV equation) that (\ref{mkdvcasimir}) and the CQ equation (\ref{htks}) are related  by the Miura transformation $\rho=u+u_{xx}$. Using this transformation the bi-Hamiltonian structure of (\ref{mkdvcasimir}) follows from the bi-Hamiltonian structure of the CQ equation derived in Section 3 and they yield
\begin{eqnarray}
{\cal D}_1\equal\partial+\partial^3\;,\quad\quad H_2=-{1\over 2}\int d\theta\,\rho^{-1}\;,\nonumber\\\noalign{\vskip 7pt}
{\cal D}_2\equal\partial\rho\partial^{-1}\!\rho\partial\;,\quad  H_1=-{1\over 8}\int d\theta\,\left(\rho^{-3}-4\rho^{-5}\rho_\theta^2\right)\;.\nonumber
\end{eqnarray}
The input of this bi-Hamiltonian structure was used in \cite{RenAlatancang} for the derivation of the bi-Hamiltonian structure of the CQ equation. In fact  they transformed ${\cal D}_1$, $H_2$ back to the CQ equation variables and the second Hamiltonian structure was obtained by a factorization of the recursion operator (\ref{urecursion}).
\section{Conclusion:}

In this paper, the transformations used by Sakovich to map the CQ  and mKdV equations into each other were shown to have a nice geometrical interpretation. Namely, they follow from a isoperimetric  curve motion in the Euclidean plane. Also, the bi-Hamiltonian structure of the system was obtained without any knowledge of its recursion operator or other known result of equations associated with the system under consideration. Finally, we have derived a hierarchy of equations, symmetries and equations associated with the CQ equation system which have appeared in the literature previously.

\section*{Acknowledgments}

We would like to thank the authors of \cite{RenAlatancang} for sending us a copy of their paper. This work was supported by CNPq (Brazil).

\end{document}